\documentclass[preprint,aps,showpacs,prb,floatfix]{revtex4}

\usepackage{graphicx}
\usepackage{dcolumn}
\usepackage{bm}

\begin{document} 


\title{Electron correlations in a C$_{20}$ fullerene cluster:\\ 
A lattice density-functional study of the Hubbard model
      } 
\author{R.~L\'opez-Sandoval$^{1,2}$ and G.~M.~Pastor$^{2}$}
\affiliation{$^1$Instituto Potosino de Investigaci\'on Cient\'{\i}fica
y Tecnol\'ogica,
Camino a la presa San Jos\'e 2055, 78216 San Luis Potos\'{\i}, M\'exico}
\affiliation{$^2$ Laboratoire de Physique Quantique, 
Centre National de la Recherche Scientifique,
Universit\'e Paul Sabatier, 31062 Toulouse, France}

\date{\today}

\begin{abstract}

The ground-state properties of C$_{20}$ fullerene clusters 
are determined in the framework of the Hubbard model by using
lattice density-functional theory (LDFT) and 
scaling approximations to the interaction-energy functional.
Results are given for the ground-state energy, kinetic and Coulomb 
energies, local magnetic moments, and charge-excitation 
gap, as a function of the Coulomb repulsion $U/t$ and for 
electron or hole doping $\delta$ close half-band filling 
($|\delta| \le 1$). The role of electron correlations is 
analyzed by comparing the LDFT results with fully unrestricted 
Hartree-Fock (UHF) calculations which take into account
possible noncollinear arrangements of the local spin-polarizations.
The consequences of the spin-density-wave symmetry breaking, 
often found in UHF, and the implications of this 
study for more complex fullerene structures are discussed.

\end{abstract}
%


%
%

%
\maketitle

\section{\label{sec:introd}
Introduction} 

The discovery of the C$_{60}$ fullerene\cite{kroto} and the 
remarkable physical and chemical properties resulting from its 
unique topology and electronic structure have motivated an extraordinary
research activity in past years.\cite{fullgen} 
One of the recent questions of interest in this field
is the possibility of producing even smaller C or Si cagelike 
clusters and to synthesize novel solids using them as building 
blocks. Thus, several experimental and theoretical studies have been 
performed in order to elucidate the complex mechanisms of formation 
of these nanostructures.\cite{cages,piskoti,cote,prinz,saito,ehlich,iqbal} 
In the particular case of C$_{20}$, 
which is expected to be the smallest fullerene, 
experiments indicate that 
the dominant species in laser-vaporization of graphite are ring 
structures, while theory yields different isomers depending 
on the calculation 
method.\cite{held,hunter,yang,hand,ragha,taylor,jones,grossman}
Hartree-Fock (HF) studies predict that the 
ring is the lowest-energy isomer, followed by the bowl 
(a substructure of the C$_{60}$ fullerene) and then 
by the dodecahedral cage.\cite{ragha,taylor,jones} In contrast, 
the local density approximation (LDA) 
to density-functional theory (DFT) yields the cage to be more stable
than the bowl and the ring, whereas in a generalized gradient 
approximation (GGA) the same ordering as HF is 
obtained.\cite{ragha,taylor,jones} 
These investigations, as well as 
quantum Monte Carlo (QMC) calculations,\cite{grossman}
indicate that non-trivial electron correlation effects play 
a central role in the structural and electronic properties
of these nanoclusters.

Recently, Prinzbach and coworkers succeeded to produce the 
cage-structured fullerene C$_{20}$ in the gas phase starting from
the perhydrogenated form C$_{20}$H$_{20}$.\cite{prinz} 
This isomer, which is not formed spontaneously in carbon 
condensation or cluster annealing processes, has a sufficiently 
long lifetime. Moreover, photoelectron spectrum (PES) 
measurements have been performed from which its cage-like 
structure has been inferred.\cite{prinz}
Theoretical studies by Saito and Miyamoto have reproduced the 
PES thereby confirming the structure assignment given in 
experiment.\cite{saito} In addition, experimental evidence 
has been provided on the oligomerization of 
C$_{20}$ fullerenes\cite{ehlich} and on the formation of a 
solid phase of C$_{20}$ dodecahedra\cite{iqbal} 
which, unlike C$_{60}$ solids, involve 
strong C-C bonds between the pentagons of neighboring units. 

Besides its experimental realization, the
topology of C$_{20}$ appears to be a particularly attractive 
physical situation for theoretical investigations of correlated 
itinerant electrons in cagelike clusters with pentagonal rings. 
Similar studies on C$_{12}$ and C$_{60}$ clusters have already 
been performed in relation to alkali-metal-doped C$_{60}$ solids 
like K$_3$C$_{60}$ and Rb$_{3}$C$_{60}$,\cite{heb} since one expects 
that the basic physics behind the superconducting and optical 
properties of these materials should be captured at the scale of 
an individual molecular constituent. The smaller fullerenes have 
attracted a special attention in this context due to the 
perspective of achieving larger superconducting transition 
temperatures than the C$_{60}$-based solids. Indeed, 
a number of theoretical investigations
predict a significant enhancement of the electron-phonon coupling 
as the size of the fullerene is reduced from C$_{60}$ to C$_{36}$, 
C$_{28}$, and finally C$_{20}$.\cite{cote,breda,saito2} 
Consequently, theoretical studies aiming at understanding the 
properties of correlated electrons in small fullerene clusters 
are of considerable interest.

Lattice models have been used 
to determine low-energy properties of 
C$_{60}$ which derive from the outermost half-filled $\pi$-electron 
cloud. In particular, the spin-density distribution on the
buckyball has been analyzed in a series of 
papers.\cite{coffeyprl,coffeyprb,ber,fal,joy,maol}
Coffey and Trugman\cite{coffeyprl}
calculated the ground-state spin configuration using a classical
antiferromagnetic (AF) Heisenberg Hamiltonian
corresponding to the strongly correlated limit of 
the Hubbard model. They found that the lowest-energy
spin structure shows a nontrivial noncollinear order which 
minimizes magnetic frustrations within each pentagonal ring, keeping strong
AF short-range order in bonds connecting
nearest neighbor (NN) pentagons.\cite{coffeyprl} 
The spin structure in C$_{60}$ was also investigated 
by means of Hubbard or Pariser-Parr-Pople Hamiltonians at 
half-band filling, taking into account on-site and inter-site Coulomb 
interactions within the unrestricted Hartree-Fock (UHF) 
approximation.\cite{ber,fal,joy,maol} 
A common result of these investigations is the presence of a
magnetic instability for a critical value $U_c$ of the on-site 
Coulomb repulsion yielding a magnetic order which resembles that 
of the classical AF Heisenberg model. 
In addition remarkable noncollinear spin arrangements and 
charge-density redistributions have been obtained in mean-field 
calculations as a function of electron and hole doping.\cite{maol}
QMC simulations on C$_{60}$ at half-band filling
and exact diagonalization studies of cagelike C$_{12}$
support the existence of non-vanishing short-range spin 
correlations.\cite{coffeyprb,moreo}
However, it should be recalled that there is no experimental
evidence for an spontaneous symmetry breaking. Instead, the 
spin-density-wave instability should be interpreted as an 
indication of fluctuating spin-spin correlations within the cluster.
It is therefore interesting to improve on the 
treatment of correlations in order to quantify the possible
effects of artificial symmetry breakings on the electronic properties.

The purpose of this paper is to investigate the ground-state properties of 
correlated electrons on a C$_{20}$ cagelike cluster by using 
the Hubbard Hamiltonian and a recently developed lattice 
density-functional theory (LDFT).\cite{ldftxcfun,ldftscfer} 
Previous applications
of this approach to 1D and 2D lattices have yield quite accurate 
ground-state properties for all band fillings and interaction regimes. A 
particularly interesting feature in the present 
context is that LDFT does not necessarily involve a symmetry 
breaking in order to account for the effects of electron correlation 
and localization in the strongly interacting 
limit.\cite{ldftscfer} Thus, it appears as an appropriate means 
of improving on UHF calculations.
Moreover, besides the aspects specifically related to C$_{20}$
and its topology, the present calculations should be of interest
as a finite cluster application of new density-functional approaches to 
lattice fermion models.\cite{otherdft} 

The body of the paper is organized as follows.
In Sec.~\ref{sec:teo} the model Hamiltonian and the main steps in the 
formulation of LDFT are briefly recalled.  
Results for the ground-state energy, kinetic and Coulomb energies,
local magnetic moments, and charge excitation gap of C$_{20}$ are 
presented and discussed in Sec.~\ref{sec:res} 
as a function of Coulomb repulsion $U/t$ and for 
electron or hole dopings $\delta$ close half-band filling 
($|\delta| \le 1$). Finally, Sec.~\ref{sec:disc} summarizes the
main conclusions and discusses some perspectives of  
extensions in view of applications to lower symmetry structures.

\section{\label{sec:teo}
Model Hamiltonian and calculation method} 

\begin{figure}
\centerline{\includegraphics[scale=1]{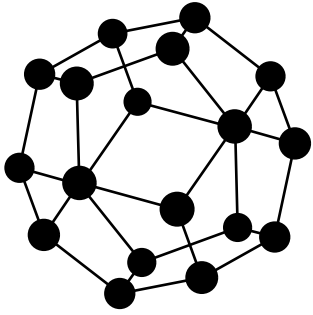} }
\caption{Illustration of the geometry of the cagelike C$_{20}$ cluster.}
\label{fig:struct} 
\end{figure}

We consider the Hubbard Hamiltonian\cite{hub}  
\begin{equation}
\label{eq:hamhub}
H = -t \sum_{\langle i,j\rangle \sigma} 
\hat c^{\dagger}_{i \sigma} \hat c_{j \sigma} +
U \sum_i  \hat n_{i \downarrow} \hat n_{i\uparrow} ,
\end{equation}
on the C$_{20}$ structure illustrated in Fig.~\ref{fig:struct}. 
In the usual notation, $\hat c_{i\sigma}^\dagger$ ($\hat c_{i\sigma}$) refers
to the creation (annihilation) operator of an electron
with spin $\sigma$ at site $i$, and 
$\hat n_{i\sigma} = \hat c_{i\sigma}^\dagger \hat c_{i\sigma}$ 
to the corresponding
number operator. The first term is the kinetic-energy operator with 
hopping integrals $t_{ij} = -t < 0$ for NNs and $t_{ij} = 0$ otherwise. 
The second term takes into account intraatomic interactions 
by means of the on-site Coulomb repulsion integral $U$. 
Given the lattice structure and the 
number of atoms $N_a = 20$, the model is characterized by the
number of electrons $N_e$, or doping $\delta = N_e -N_a$, and by the
ratio $U/t$.  

In order to determine the ground-state properties we consider
a density-functional approach to electron correlations in a 
lattice, in which the fundamental variable is 
the single-particle density matrix $\gamma_{ij}$.\cite{ldftscfer}
The ground-state energy $E_{gs}$ and density matrix $\gamma_{ij}^{gs}$
are determined by minimizing the energy functional  
\begin{equation}
\label{eq:E}
E[\gamma] = E_K[\gamma] + W [\gamma]
\end{equation}
with respect to $\gamma_{ij}$. 
The first term in Eq.~(\ref{eq:E}) is the kinetic energy
\begin{equation}
\label{eq:EK}
E_K[\gamma] = 
\sum_{ij} t_{ij} \gamma_{ij} \; 
\end{equation}
associated with the electronic motion in the lattice.
The second term is the interaction-energy functional,
formally given by\cite{levy}
\begin{equation}
\label{eq:Wex}
W[\gamma] = 
{\min_{\Psi\to\gamma}} \left[ U \sum_i  \langle \Psi [\gamma] |
\hat n_{i\uparrow} \hat n_{i\downarrow}|
\Psi [\gamma] \rangle  \right] \; ,
\end{equation}
where the constrained minimization runs over all 
$N$-particle states $| \Psi [\gamma] \rangle$ 
that satisfy 
\begin{equation}
\label{eq:rep}
\langle \Psi [\gamma] | \; 
\sum_\sigma \hat c_{i \sigma }^{\dagger}\hat c_{j \sigma} \; 
|\Psi [\gamma] \rangle = \gamma_{ij} 
\end{equation}
for all $i$ and $j$. Thus, $W$ represents the 
minimum value of the interaction energy compatible with a given
degree of electron delocalization or  
density matrix $\gamma_{ij}$. The universal functional $W[\gamma]$, 
valid for all lattice structures and hybridizations, 
can be considerably simplified if the hopping integrals are 
short ranged. For example, if only NN hoppings are considered 
as in the present case, the kinetic energy $E_K$ is 
independent of the density-matrix elements between sites that are not NNs. 
Therefore, the constrained search in Eq.~(\ref{eq:Wex}) may be restricted 
to the $| \Psi [\gamma] \rangle$ that satisfy Eq.~(\ref{eq:rep})
only for $i = j$ and for NN $ij$. Moreover, for highly symmetric 
clusters like C$_{20}$, one has the same equilibrium value of $\gamma_{ij}$ 
for all NN pairs $ij$, and $\gamma_{ii} = n = N_e/N_a$ for all sites $i$. 
The interaction energy can then be regarded as a simple function 
$W(\gamma_{ij})$ of the density-matrix element between NNs. 
However, notice that restricting the minimization constraints 
in Eqs.~(\ref{eq:Wex}) and (\ref{eq:rep}) 
to NN $\gamma_{ij}$ also implies that $W$ loses its 
universal character, since the NN map and the resulting dependence 
of $W$ on $\gamma_{ij}$ are in principle different for different 
lattice structures.\cite{ldftxcfun,ldftscfer}

Two previously investigated scaling approximations to the interaction 
energy $W$ of the Hubbard model shall be used for the 
calculations.\cite{ldftscfer,ldft4order} 
In the first one the functional dependence 
is derived from the exact solution of Eq.~(\ref{eq:Wex}) for the 
Hubbard dimer. It is given by\cite{ldftscfer}
\begin{equation}
W^{(2)} = E_{\rm HF} 
\left( 1 - \sqrt{1 - g_{ij}^2 } \right) \; ,
\label{eq:W2}
\end{equation}
where $E_{\rm HF} = N_a U n^2/4$ refers to the Hartree-Fock energy
and $g_{ij} = (\gamma_{ij}     - \gamma_{ij}^{\infty}) / 
              (\gamma_{ij}^{0} - \gamma_{ij}^{\infty})$
measures the degree of electron correlation in a NN bond $ij$.
Here, $\gamma_{ij}^0 > 0$ stands for the largest possible 
value of the bond order $\gamma_{ij}$ for a given 
$N_a$, band filling $n=N_e/N_a$, and lattice structure. It 
represents the maximum degree of electron delocalization 
regardless of correlations. $\gamma_{ij}^\infty$ refers to the strongly 
correlated limit of $\gamma_{ij}$, i.e., to the largest NN bond 
order that can be achieved under the constraint of vanishing $W$.
For half-band filling $\gamma_{ij}^\infty = 0$, 
while for $n \not= 1$, $\gamma_{ij}^\infty > 0$.\cite{nonbip}
Thus, $g_{ij} = 1$ in an uncorrelated state, and $g_{ij} = 0$ in the 
strongly correlated limit corresponding, for example,  
to a fully localized or fully spin-polarized state (Nagaoka state).
Physically, the scaling hypothesis
underlying Eq.~(\ref{eq:W2}) means that the relative change 
in $W$ associated with a given change in $g_{12}$
is considered as independent of the system under study. 
Exact numerical studies of the functional dependence of $W$
have shown that this is a good approximation in a wide variety 
of 1D, 2D, and 3D lattices and band fillings.\cite{ldftxcfun}
Notice that $E_{\rm HF}$, $\gamma_{12}^{\infty}$, and $\gamma_{12}^0$ 
are system specific. In practice, $\gamma_{12}^{\infty}$  
is approximated by the ferromagnetic fully-polarized 
$\gamma_{12}^{\rm FM}$ which can be obtained, as $E_{\rm HF}$ 
and $\gamma_{12}^0$, from the single-particle electronic structure.

The dimer approximation to $W$ is very appealing since it
combines remarkable simplicity and good accuracy.\cite{ldftscfer,ldftdim1D} 
Nevertheless, it also presents some limitations particularly
in the limit of strong correlations at half-band filling. 
For $N_e = N_a$ and small $\gamma_{ij}$, $W^{(2)}$ can be expanded as 
$W^{(2)} = (1/8) \alpha_2 U \gamma_{ij}^2 + {\cal O}(\gamma_{ij}^4)$
with $\alpha_2 = (\gamma_{ij}^0)^{-2}$ 
(e.g., $\alpha_2 = 2.92$ for a pentagonal ring and $\alpha_2 = 4.16$
for C$_{20}$). The exact functional $W$ shows the same behavior
but usually with a somewhat larger coefficient $\alpha_{ex}$
(e.g., $\alpha_{ex} = 3.21$ for a pentagonal ring and $\alpha_2 = 5.22$
for C$_{20}$).\cite{exact-coef} These discrepancies have direct 
consequences on the resulting ground-state properties 
for $U/t\gg 1$, since in this limit
$\gamma_{ij}^{gs} \simeq (4z/\alpha)(t/U)$ and
$E_{gs} \simeq -(2z^2/\alpha)(t^2/U)$,
where $z$ is the local coordination number ($E_K = -zt\gamma_{ij}$). 
In order to improve the accuracy a more flexible 
approximation has been proposed which is also
based on the scaling properties of $W$, and which recovers the
exact dependence on $\gamma_{ij}$ for $\gamma_{ij} \to 0$.
In this case the interaction energy includes a fourth-order 
term in $g_{ij}$ and is given by
\begin{equation}
W^{(4)} = E_{\rm HF} 
\left( 1 - \sqrt{1 - \kappa g_{ij}^2 + (\kappa -1) g_{ij}^4} \right) \; ,
\label{eq:W4}
\end{equation}
where $\kappa = \alpha_{ex} / \alpha_2$. In a pentagonal ring 
$\kappa = 1.10$, while in the C$_{20}$ cluster $\kappa = 1.26$. 
These values should be compared with the dimer result $\kappa = 1$, 
for which Eq.~(\ref{eq:W4}) reduces to Eq.~(\ref{eq:W2}). 
Thus, the 4th-order term appears as a moderate correction 
to the dimer or 2nd-order approximation ($g_{ij}^2 \le 1$). 
Previous applications to 1D and 2D systems
have shown that Eq.~(\ref{eq:W4}) provides a systematic 
improvement on the ground-state properties for all $U/t$.\cite{ldft4order}  
In the following section, Eqs.~(\ref{eq:W2}) and (\ref{eq:W4}) are applied 
to determine several electronic properties of the C$_{20}$ Hubbard cluster 
in the framework of LDFT.

\section{\label{sec:res}
Results}  

\begin{figure}
\includegraphics[scale=.4]{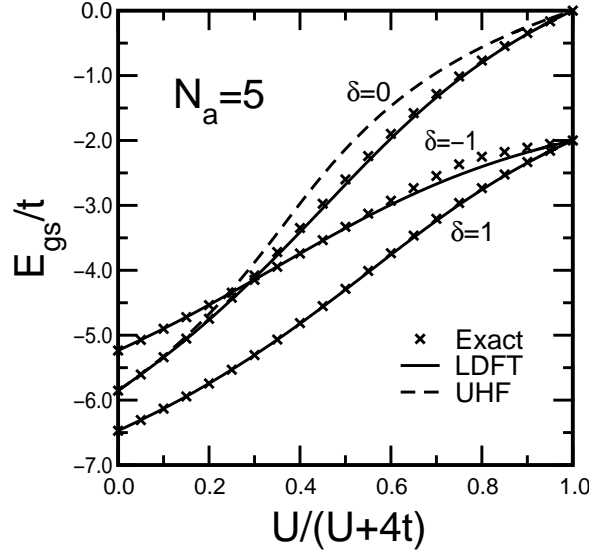}
\caption{
Ground-state energy $E_{gs}$ of the Hubbard model on a pentagonal ring as 
a function of Coulomb interaction $U/t$ for different numbers of electrons 
$N_e = N_a + \delta$ close to half-band filling ($N_a =5$). 
For $\delta = 1$, results for $E_{gs} - U$ are shown.
The solid curves correspond to lattice density-functional theory
(LDFT), the crosses to exact numerical diagonalizations,
and the dashed curve for $\delta = 0$ to unrestricted Hartree-Fock
(UHF) calculations including noncollinear spin arrangements.
        }  
\label{fig:Egsrg5}
\end{figure}

The model is characterized by the dimensionless parameter $U/t$ 
and by the doping $\delta = N_e -N_a$. We consider the complete 
range of repulsive interactions ($U\ge 0$) and single electron or hole
dopings ($|\delta| \le 1$).
Commonly accepted values of $U/t$ for the $\pi$-electrons
in carbon fullerenes correspond to the intermediate range 
$2\le U/t\le 5$.\cite{ber,fal,joy,maol} 
Before discussing the results for the C$_{20}$ cluster 
it is useful to consider a single pentagonal ring as a preliminary test 
on the accuracy of the method, since the pentagon 
constitutes the basic building 
block of the C$_{20}$ structure, and since it is small enough so that 
exact diagonalizations can be easily performed. 
Fig.~\ref{fig:Egsrg5} shows results for the ground-state energy
$E_{gs}$ of the $N_a = 5$ ring as a function of $U/t$. In the uncorrelated
limit $E_{gs}$ decreases monotonously with increasing number of 
electrons $N_e = N_a +\delta\le 6$ since the three lowest 
single-particle eigenvalues of the ring are negative. On the other side,
for large $U/t$, $E_{gs}$ vanishes at half-band filling 
while for negative (positive) doping $E_{gs}$ ($E_{gs} - U \delta$) remains 
negative. In fact, for $\delta = 0$
the suppression of charge fluctuations always implies electron 
localization, whereas for $|\delta | = 1$ a finite kinetic energy persists,
even for $U/t \gg 1$, due to the delocalization 
of the extra electron or hole.
The LDFT calculations reproduce remarkably well the crossover between
weak and strong correlations as given by the exact diagonalizations.
Appreciable discrepancies are found only for $\delta = -1$
and $U/t \ge 6$, in which case the largest relative error amounts to  
$|E_{gs}^{LDFT}- E_{gs}^{ex}| / |E_{gs}^{ex}| \simeq  0.07$ 
for $U/t \simeq 12$. In the other cases ($\delta = 0$ or $1$) 
the LDFT results are practically indistinguishable from the exact ones.

In order to quantify the effects of electron correlations we
have also performed calculations by using the fully unrestricted
Hartree-Fock (UHF) approximation which corresponds to the most general 
single-Slater-determinant wave-function and which allows for 
noncollinear site-dependent spin polarizations 
$\langle \vec S_i \rangle$.\cite{maol} 
Figure~\ref{fig:Egsrg5} shows the UHF results for $E_{gs}$ 
of the pentagonal ring at half-band filling   ($\delta = 0$). 
For $U/t < U_c \simeq 0.27$ the only 
solution to the self-consistent equations is non-magnetic
(i.e., $\langle \vec S_i \rangle = 0$ for all $i$), while for 
$U/t > U_c$ local magnetic moments $\langle \vec S_i \rangle$
set in. These increase monotonously with $U/t$ reaching 
saturation in the limit of $U/t \to\infty$.
Concerning the magnetic order one observes that the 
$\langle \vec S_i \rangle$ are all coplanar and that 
the angle between NN moments is $4\pi /5$ for all $U/t > U_c$. 
This amounts to split one parallel-spin frustration among five
bonds which corresponds, as expected for half-band filling, 
to the solution of a classical 
Heisenberg model with antiferromagnetic NN interactions.
One observes that UHF reproduces the $U/t$ dependence of 
$E_{gs}$ qualitatively well, including the strongly
correlated limit where $E_{gs} \to 0$. This is achieved by localizing 
the electrons through the formation of increasingly large local 
magnetic moments as $U/t$ increases. Nevertheless, significant 
quantitative discrepancies with the exact numerical solution are 
observed already for $U/t \ge 2$, which 
reflect the importance of electron-correlation effects. 
These can be regarded in part as the result of fluctuations between 
different spin configurations which cannot be recovered 
in a single-determinant state.\cite{bois}
In contrast, LDFT gives a much better description of the ground-state, 
particularly in the intermediate and strong interacting regimes. 
UHF overestimates $E_{gs}$ appreciably, while the differences between 
exact and LDFT results are very small. 

\begin{figure}
\includegraphics[scale=.4]{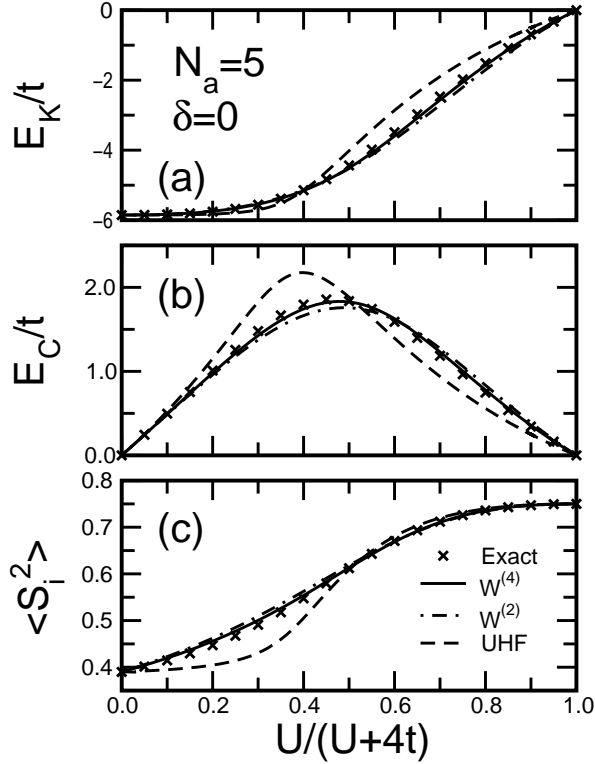}
\caption{
(a) Kinetic energy $E_{K}$, (b) Coulomb energy $E_C$, and 
(c) local magnetic moments $\langle S_i^2 \rangle$ of the 
Hubbard model on a pentagonal ring at half-band filling
($N_e = N_a = 5$). 
The dashed-dotted and full curves correspond to lattice 
density-functional theory
(LDFT) using, respectively, Eqs.~(\ref{eq:W2}) and (\ref{eq:W4}) 
for the interaction-energy functional. 
The dashed curves refer to the unrestricted Hartree-Fock
(UHF) calculations including noncollinear spin arrangements,
and the crosses to exact numerical diagonalizations.
        }
\label{fig:EkEcrg5}
\end{figure}

More detailed information on the correlation effects 
and on the accuracy of the calculations is obtained from the kinetic 
energy $E_K$, Coulomb energy $E_C$, and local square magnetic moments 
$\langle S^2_i\rangle$ presented in Fig.~\ref{fig:EkEcrg5}. 
LDFT accurately reproduces the exact numerical 
solution, showing that the very good results obtained for $E_{gs}$ 
are not the consequence of important compensations of errors.
Moreover, the differences between 2nd-order and 4th-order 
approximations to $W$ [see Eqs.~(\ref{eq:W2}) and (\ref{eq:W4})]
are quite small, typically of the order of the differences 
between the 4th-order calculations and the exact results.
Nevertheless, one may observe that the interaction-energy 
functional $W^{(4)}$ provides a consistent improvement with 
respect to $W^{(2)}$. In fact taking into account the 
4th-order term yields a small increase (reduction) of $E_C$ 
for $U/t < 4$ ($U /t > 4$) as well as a small reduction of 
$|E_K|$ for $U /t > 4$ which bring the outcome of LDFT systematically 
closer to the exact solution.\cite{foot-acc} 

In contrast to LDFT the UHF calculations do not provide a correct 
description of the kinetic and Coulomb energies separately.
On the one side, for $U/t < 4$, UHF overestimates the Coulomb 
energy $E_C$ due to an underestimation of the local moments 
$\langle S_i^2 \rangle$ (see Fig.~\ref{fig:EkEcrg5}). On the other 
side, for $U/t > 4$, both $E_C$ and $|E_K|$ are appreciably
underestimated. The inaccuracies in $E_C$ and $|E_K|$ for $U/t > 4$
tend to cancel each other, which improves to some extent the 
UHF result for $E_{gs}$ (see Fig.~\ref{fig:Egsrg5}). Notice that 
UHF reproduces quite accurately the local moments for $U/t >4$
(see Fig.~\ref{fig:EkEcrg5}). However, a single-determinant state 
fails to incorporate correlations, and in particular the fluctuations 
between different equivalent spin configurations, which should 
restore the broken spin 
symmetry and which are often important in finite systems.\cite{bois}
This reflects the limitations of the broken symmetry 
UHF solutions, even in the most general noncollinear case.\cite{maol} 
Notice that in LDFT no artificial symmetry breaking is required 
in order to account for correlation-induced localizations
and the resulting effects on the ground-state properties.
The present LDFT formalism with the dimer or 4th-order scaling 
approximations to interaction-energy functional 
describes correctly the electronic correlations in a pentagonal 
Hubbard ring. It can be therefore expected that the method 
should remain accurate in more complex systems with five-fold 
symmetry such as the C$_{20}$ cluster to be discussed below.
Moreover, comparison with exact results also shows that the differences 
between UHF and LDFT results for $E_{gs}$, $E_K$, and $E_C$ are
a good quantitative estimation of the correlation energies 
and their $U/t$ dependence.

\begin{figure}
\includegraphics[scale=.35]{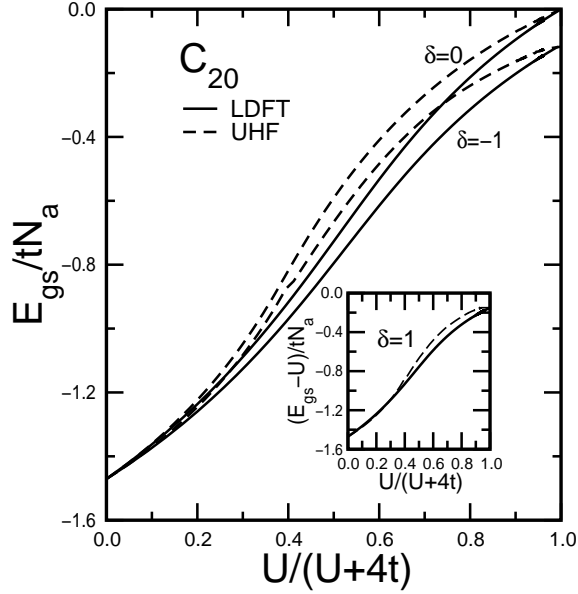}
\caption{
Ground-state energy $E_{gs}$ of the Hubbard model on the cagelike 
C$_{20}$ cluster illustrated in Fig.~\protect\ref{fig:struct}. 
Results are given as a function of Coulomb interaction 
$U/t$ for different numbers of electrons $N_e = N_a + \delta$ 
close to half-band filling ($N_a =20$). In the inset 
$E_{gs} - U$ is shown for $\delta = 1$. 
The solid curves correspond to lattice density-functional theory
(LDFT) using Eq.~(\protect\ref{eq:W4}) for the interaction-energy 
functional. The dashed curves refer to
unrestricted Hartree-Fock (UHF) calculations including 
noncollinear spin arrangements.
        }  
\label{fig:EgsC20}
\end{figure} 
\begin{figure}
\includegraphics[scale=.4]{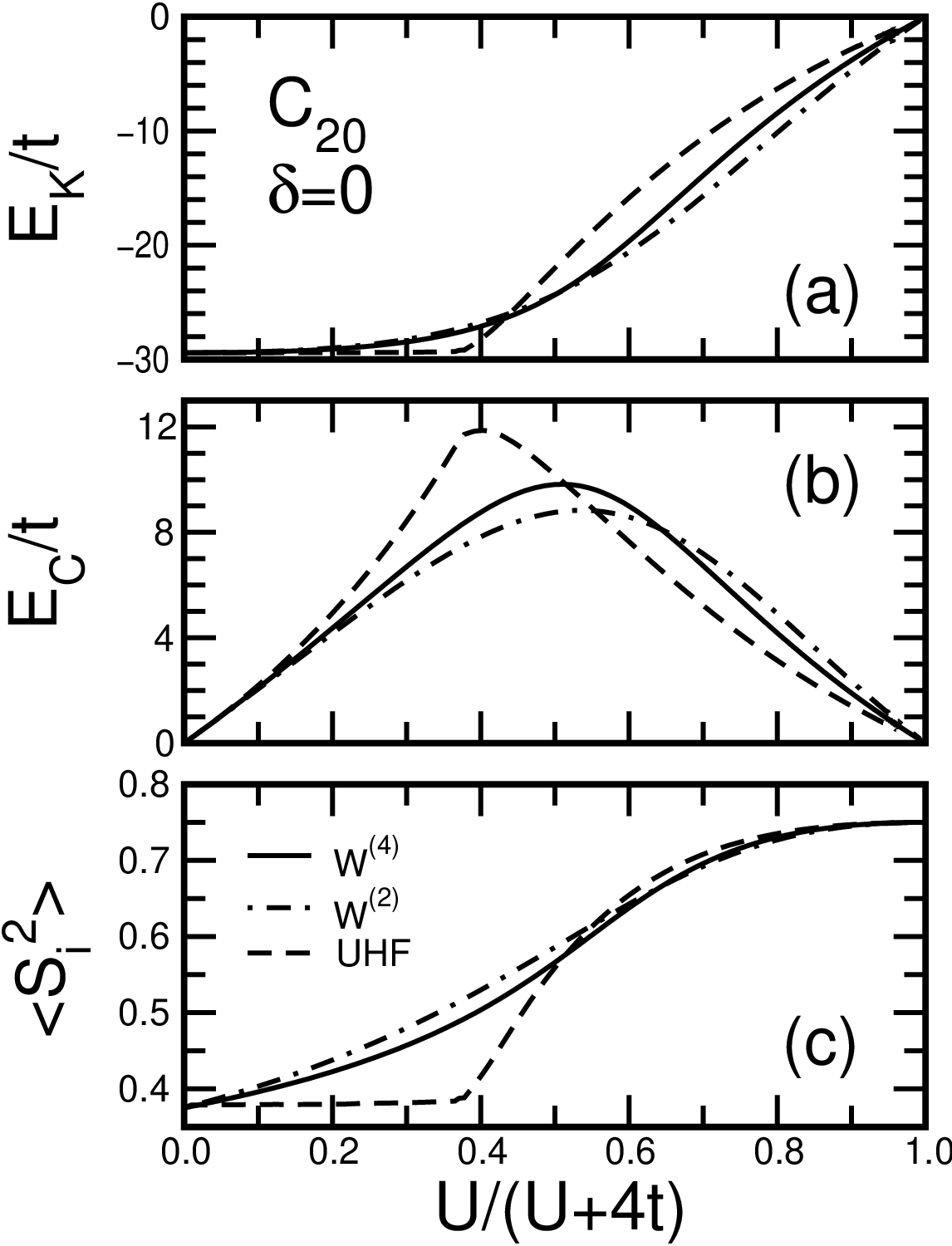}
\caption{
(a) Kinetic energy $E_{K}$, (b) Coulomb energy $E_C$, and 
(c) local magnetic moments $\langle S_i^2 \rangle$ of the cagelike
C$_{20}$ Hubbard cluster at half-band filling ($\delta = 0$). 
The dashed-dotted and full curves correspond to lattice 
density-functional theory
(LDFT) using, respectively, Eqs.~(\ref{eq:W2}) and (\ref{eq:W4}) 
for the interaction-energy functional. 
The dashed curves refer to unrestricted Hartree-Fock
(UHF) calculations including noncollinear spin arrangements.
} 
\label{fig:EkEcC20}
\end{figure}

In Figs.~\ref{fig:EgsC20} and \ref{fig:EkEcC20} LDFT and 
UHF results are given for several ground-state
properties of the Hubbard model on the cagelike C$_{20}$ 
cluster illustrated in Fig.~\ref{fig:struct}. In the uncorrelated limit 
the ground-state energies $E_{gs}$ are the same for $|\delta| \le 1$, 
a consequence of the degeneracy of the single-particle spectrum.
As $U/t$ increases, $E_{gs}$ increases monotonously with a slope 
$\partial E_{gs} / \partial U = 
 \sum_i \langle n_{i\uparrow} n_{i\downarrow}\rangle >0$. 
Since the average number of double occupations is larger
for larger number of electrons the degeneracy with respect 
to $\delta$ is removed at finite $U/t$. On the other side, 
in the strongly correlated limit, $E_{gs}$ vanishes at half-band 
filling, while for $\delta = -1$ ($\delta = 1$) 
$E_{gs}$ ($E_{gs} -U$) remains negative. The $U/t$ dependence
of $E_{gs}$ in C$_{20}$ follows similar trends as in 
the pentagonal ring. The UHF calculations reproduce the exact 
limits for $U/t = 0$ and $U/t = \infty$, and agree qualitatively 
with LDFT for finite values of the Coulomb repulsion. However, the
quantitative differences can be quite significant particularly
for intermediate and large $U/t$ (e.g., 26\% 
for $U/t \simeq 36$).

\begin{figure}
\includegraphics[scale=.5]{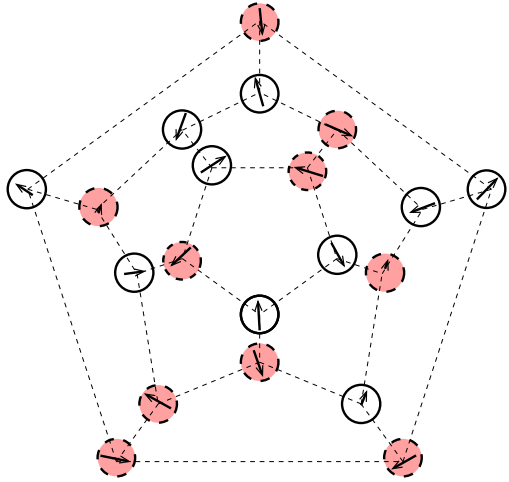}
\caption{
Illustration of the distribution of the local spin polarizations 
$\langle\vec S_i\rangle$ in an undoped cagelike C$_{20}$ cluster 
as obtained by using the Hubbard model and the unrestricted 
Hartree-Fock approximation ($\delta = 0$). Due to spin-rotational
invariance only the relative orientations of $\langle\vec S_i\rangle$ 
are physically meaningful. The three-dimensional 
cluster structure (see Fig.~\protect\ref{fig:struct}) is mapped 
onto a plane in order to ease the visualization. The radius of the 
circle on each atomic site $i$ is proportional to 
$\vert \langle \vec S_i \rangle\vert$.  The arrows represent  
the projection of $\langle \vec S_i\rangle$
onto the $xy$ plane which is the plane containing the 
outermost pentagon. Shaded circles (open circles) indicate that the 
perpendicular component $\langle S_i^z\rangle$ is positive (negative).
        }  
\label{fig:c20sdw}
\end{figure}

The effects of correlations are more significant 
in the kinetic energy $E_K$, Coulomb energy $E_C$, and local 
square moment $\langle S^2_i\rangle$ shown in Fig.~\ref{fig:EkEcC20}
for $\delta = 0$. The differences 
between UHF and LDFT calculations for $E_K$ and $E_C$ are similar 
to the corresponding differences between UHF and exact results for 
the pentagonal ring and can be qualitatively understood in similar 
terms. In the case of C$_{20}$ one may notice that the 
UHF results for the local square moment $\langle S^2_i\rangle$
are the same as in the uncorrelated limit for $U/t < 2.2$. Therefore,
they are significantly underestimated. For stronger Coulomb
repulsions ($U/t > 2.2$) $\langle S^2_i\rangle$ increases rapidly 
reaching values that are
very close to the LDFT results for $U/t \simeq 4$
(see Fig.~\ref{fig:EkEcC20}). This behavior is related 
to the $U/t$ dependence of the self-consistent solution of 
the UHF equations. For $0\le U/t \le 3.1$ the UHF charge 
distribution is not uniform due to the 
four-fold degeneracy of the single-particle spectrum 
at the Fermi energy. Moreover, for $U/t < 2.2$
the UHF ground state shows a weakly ferromagnetic solution 
with very small local spin polarizations. In this way 
advantage is taken from the single-particle degeneracy 
in order to reduce the Coulomb-repulsion energy without 
increasing significantly the kinetic energy.  
For $U/t > 2.2$ the system starts to develop an AF
spin-density wave which coexists with a weak charge-density 
wave for $2.2 < U/t < 3.1$. Only for $U/t > 3.1$ the AF-like 
magnetic order valid for strong interactions is definitely 
settled and the charge distribution becomes then uniform 
throughout the cluster. An illustration of the UHF magnetic order
in the large $U/t$ regime is shown in Fig.~\ref{fig:c20sdw}.
Notice the noncollinearity of the local spin polarizations 
$\langle\vec S_i\rangle$, which form angles of about 138 
degrees between NNs. This value is smaller than the angle
between NN $\langle\vec S_i\rangle$ in an isolated pentagonal ring 
($144$ degrees) because of frustration effects between the pentagons 
in the C$_{20}$ topology (see Figs.~\ref{fig:struct} and 
\ref{fig:c20sdw}). Further increase of the Coulomb repulsion 
beyond $U/t = 3.1$ only results in an increase of the size 
of the local moments keeping the magnetic order unchanged.

\begin{figure}
\includegraphics[scale=.4]{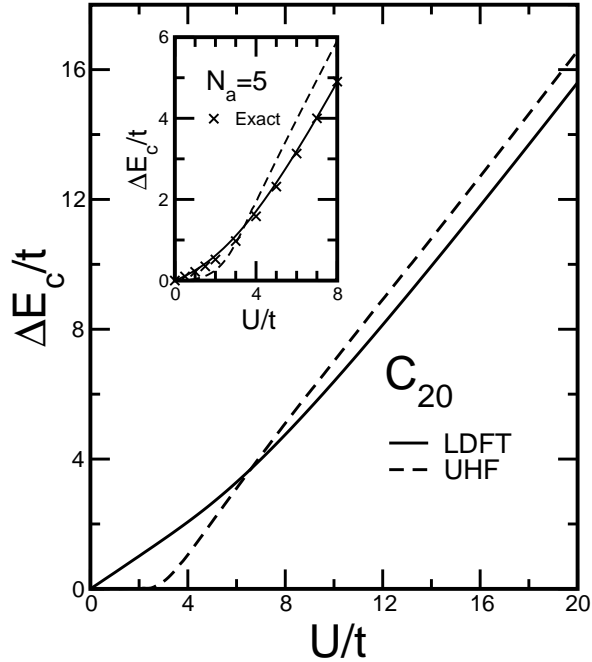}
\caption{
Charge excitation gap $\Delta E_c$ of the Hubbard model on the 
cagelike C$_{20}$ cluster as a function of Coulomb interaction $U/t$.
The solid (dashed) curves correspond to LDFT (UHF).
In the inset figure results are given for the pentagonal Hubbard ring. 
Here the crosses refer to exact diagonalizations.} 
\label{fig:gaps} 
\end{figure}

The charge excitation gap is given by 
\begin{equation}
\Delta E_c = E_{gs}(N_e + 1) +  E_{gs}(N_e - 1) - 2 E_{gs}(N_e) \; ,
\end{equation}
where $E_{gs}$ refers to ground-state energy. It is a property 
of considerable interest since it measures the low-energy 
excitations associated with changes in the number of electron 
$N_e$ and is thus very sensitive to electron correlation effects. 
In Fig.~\ref{fig:gaps} results are shown for 
$\Delta E_c$ of the C$_{20}$ Hubbard cluster at half-band filling
($N_e = N_a$). The corresponding calculations for the pentagonal 
ring are reported in the inset for the sake of comparison.
At half-band filling $\Delta E_c = 0$ for $U=0$, despite the finite size,
due to the degeneracy of the single-particle spectrum in both the 
ring and C$_{20}$. $\Delta E_c$ increases with increasing $U/t$ 
approaching the limit $\Delta E \to (U-w_b)$ for 
$U/t \to \infty$, where $w_b$ is the bottom of
single-particle spectrum of the cluster ($w_b = 2t$ for the ring 
and $w_b = 3t$ for C$_{20}$). 
The LDFT results and the exact calculations for $N_a = 5$
yield a linear dependence of $\Delta E_c$ as a function $U/t$ for 
$U/t\ll 1$. This implies, in particular for C$_{20}$, that
$\Delta E_c >0$ for arbitrary small $U/t$, in contrast to 
the UHF result. Comparison with exact diagonalizations 
for $N_a = 5$ indicates that the LDFT gap is not overestimated.
The UHF calculations fail to reproduce the correct $U/t$ 
dependence of $\Delta E_c$ showing once more the importance of 
correlations. Notice that the differences between UHF and 
exact or LDFT results are qualitatively similar for C$_{20}$ 
and for the pentagonal ring, which indicates that the contribution of 
correlations to $\Delta E_c$ are comparable in both cases.

\section{\label{sec:disc}
Conclusion}  

A density-functional approach to lattice-fermion models has been 
applied to study the ground-state properties of the 
Hubbard Hamiltonian on a cagelike C$_{20}$ cluster. The ground-state
energy, the kinetic and Coulomb contributions, and the charge 
excitation gap have been determined as a function of $U/t$ for
dopings $\delta$ close to half band filling. The importance of 
electron correlations has been quantified by performing
noncollinear Hartree-Fock calculations. Comparisons with exact 
diagonalizations and with UHF in the case of a pentagonal ring 
demonstrate the ability of LDFT to describe subtle correlation 
effects in a very simple and systematic way. It is 
therefore expected that LDFT should be an efficient tool for 
investigating the properties of strongly correlated fermions in 
more complex clusters and nanostructures, which are otherwise 
very difficult to tackle with other accurate methods.

The present LDFT study of clusters encourages the 
development of more flexible interaction-energy functionals 
in order to determine the properties of finite systems with 
lower symmetry. As a first step it would be interesting to 
extend the calculations to fullerene structures having 
different types of bonds, for example C$_{60}$. 
In these cases the functional dependence of $W[\gamma_{ij}]$ 
is more complex and one has to consider at least two different 
NN density-matrix elements, within and between pentagons, 
even if one assumes that the hopping integrals are the same. 
The symmetry of these bonds is actually
different despite the fact that all sites are equivalent.
The interaction-energy functional should be generalized
following previous works on dimerized 1D chains.\cite{ldftdim1D} 
The domain of representability of $\gamma_{ij}$ and
the accuracy of the scaling hypothesis could be explicitly 
tested for smaller similar clusters like C$_{12}$
by solving Levy's constrained search.\cite{levy,ldftxcfun} 
Furthermore, the functional dependence of $W$ for low-symmetry clusters
deserves to be explored by taking into account charge transfers 
and eventual redistributions of the spin density 
between non-equivalent sites. This should open the way 
to investigations of correlation effects in a variety 
of complex physical situations such as metal clusters, 
disordered systems, and low-dimensional nanostructures.

\begin{acknowledgments}

This work has been supported by CONACyT Mexico (Grant No.\ J-41452 
and Millennium Initiative W-8001) and by the EU GROWTH project AMMARE 
(Contract No.\ G5RD-CT-2001-00478). 
Computer resources were provided by IDRIS (CNRS, France).  

\end{acknowledgments}


\begin{thebibliography}{}


\bibitem{kroto} 
H. W. Kroto, J. R. Heath, S. C. O'Brien, R. F. Curl, and R. E. Smalley, 
Nature (London) {\bf 318}, 162 (1985);
W. Kr\"atschmer, L. D. Lamb, K. Fostiropoulos, and D. R. Huffman,
Nature (London) {\bf 347}, 354 (1990).

\bibitem{fullgen}
Y. H. Kim, I. H. Lee, K. J. Chang, and S. Lee, 
Phys. Rev. Lett. {\bf 90}, 065501 (2003); 
Y. Zhao, B. Y. Yakobson and R. E. Smalley, 
Phys. Rev. Lett. {\bf 88}, 185501 (2002); 
D. Herschbach, 
Rev. Mod. Phys. {\bf 71}, S411 (1999); 
R. E. Smalley, 
Rev. Mod. Phys. {\bf 69}, 723 (1997).

\bibitem{cages} 
U. R\"othlisberger, W. Andreoni  and M. Parrinello 
Phys. Rev. Lett. {\bf 72}, 665 (1994);
L. Mitas, J. C. Grossman, I. Stich and J. Tobik
Phys. Rev. Lett. {\bf 84}, 1479 (2000); 
Q. Sun, Q. Wang, P. Jena, B. K. Rao and Y. Kawazoe, 
Phys. Rev. Lett. {\bf 90}, 135503 (2003);
D. Conn\'etable, {\em et al.}, 
Phys. Rev. Lett. {\bf 91}, 247001 (2003);
M.N. Huda and A.K. Ray, 
Eur. Phys. J. D {\bf 31}, 63 (2004).  

\bibitem{piskoti}
C. Piskoti, J. Yarger and A. Zettl,
Nature {\bf 393}, 771 (1998).

\bibitem{cote}
M. Cot\'e, J. C. Grossman, M. L. Cohen, and S.G. Louie, 
Phys. Rev. Lett. {\bf 81}, 697 (1998). 

\bibitem{prinz}
H. Prinzbach, {\em et al.}, 
Nature {\bf 407}, 60 (2000).

\bibitem{saito}
M. Saito and Y. Miyamoto,
Phys.\ Rev.\ Lett.\ {\bf 87}, 035503 (2001).

\bibitem{ehlich}
R. Ehlich, P. Landenberger and H. Prinzbach,
J. Chem. Phys. {\bf 155}, 5830 (2001).

\bibitem{iqbal}
Z. Iqbal, {\em et al.}, 
Eur. J. Phys. B {\bf 31}, 509 (2003). 

\bibitem{held} 
G. von Helden, M. T. Hsu, N. G. Gotts, P. R. Kemper, and M. T. Bowers, 
Chem. Phys. Lett. {\bf 204}, 15 (1993).

\bibitem{hunter} 
J. N. Hunter, J. L. Fey, and M. F. Jarrold, 
J. Phys. Chem. {\bf 97}, 3460 (1993); 
Science {\bf 260}, 784 (1993)

\bibitem{yang} 
S. Yang, K. J. Taylor, M. J. Craycraft, J. Conceicao, C. L. Pettiete, 
O. Cheshnovsky, and R. E. Smaley, 
Chem. Phys. Lett. {\bf 144}, 431 (1988). 

\bibitem{hand} 
H. Handschuh, G. Gantef\"or, B. Kessler, P. S. Bechthold, and W. Eberhardt, 
Phys. Rev. Lett. {\bf 74}, 1095, (1995).  

\bibitem{ragha} 
K. Raghavachari, D. L. Striut, G. K. Odom, G. E. Scuseria, J. A. Pople, 
B. G. Johnson, and P. M. W. Gill, 
Chem. Phys. Lett. {\bf 214}, 357 (1993)

\bibitem{taylor}
P. R. Taylor, E. Bylaska, J. H. Weare, and R. Kawai, 
Chem. Phys. Lett. {\bf 235}, 538 (1995) 

\bibitem{jones}
R.O. Jones and G. Seifert, 
Phys.\ Rev.\ Lett.\ {\bf 79}, 443 (1997);
G. Galli, F. Gygi, and J.-Christophe Golaz,
Phys. Rev. B {\bf 57}, 1860 (1998).

\bibitem{grossman} 
J. C. Grossman, L. Mitas and K. Raghavachari, 
Phys. Rev. Lett. {\bf 75}, 3870 (1995).

\bibitem{heb} 
A. F. Hebard {\em et al.}, 
Nature {\bf 350}, 600 (1991); 
K. Holczer {\em et al.}, 
Science {\bf 252}, 1154 (1991);
Kroto {\em et al.}, 
Nature {\bf 318}, 162 (1985).

\bibitem{breda}
N. Breda {\em et al.}, 
Phys. Rev. B {\bf 62}, 130 (2000).

\bibitem{saito2}
M. Saito and Y. Miyamoto,
Phys.\ Rev.\ B {\bf 65}, 165434 (2002);
J. Lu {\em et al.}, 
Phys. Rev. B {\bf 67}, 125415 (2003).

\bibitem{coffeyprl} 
D. Coffey and S. A. Trugman, 
Phys. Rev. Lett. {\bf 69}, 176 (1992).

\bibitem{coffeyprb} 
D. Coffey and S. A. Trugman, 
Phys. Rev. B {\bf 46}, 12717 (1992).

\bibitem{ber} 
L. Bergomi, J. P. Blaizot, and Th. Jolicoeur and E. Dagotto,
Phys. Rev. B {\bf 47}, R5539 (1993).

\bibitem{fal} 
F. Willaime and L. M. Falicov, 
J. Chem. Phys. {\bf 98}, 6369 (1993). 

\bibitem{joy} 
P. Joyes, R. J. Tarento, 
Phys. Rev. B {\bf 45}, 12077 (1992); 
P. Joyes, R. J. Tarento, L. Bergomi,
Phys. Rev. B {\bf 48}, 4855 (1993).

\bibitem{maol} 
M. A. Ojeda, J. Dorantes-D\'avila and G. M. Pastor, 
Phys. Rev. B {\bf 60}, 6121 (1999); 
{\it ibid.} {\bf 60} 9122 (1999). 

\bibitem{moreo} 
R. T. Scalettar, A. Moreo, E. Dagotto, L. Bergomi, T. Jolicoeur, 
and H. Monien, 
Phys. Rev. B {\bf 47}, 12316 (1993).

\bibitem{ldftxcfun}
R. L\'opez-Sandoval and G.M. Pastor, 
Phys. Rev. B {\bf 61}, 1764 (2000).

\bibitem{ldftscfer}
R. L\'opez-Sandoval and G.M. Pastor, 
Phys. Rev. B {\bf 66}, 155118 (2002). 

\bibitem{otherdft}
For alternative density-functional approaches to lattice-fermion
models see, for instance, 
A.E.\ Carlsson, 
Phys.\ Rev.\ B {\bf 56}, 12058 (1997); 
R.G.\ Hennig and A.E.\ Carlsson, 
{\it ibid.} {\bf 63}, 115116 (2001);
N.A.\ Lima, M.F.\ Silva, L.N.\ Oliveira and K.\ Capelle,
Phys.\ Rev.\ Lett.\ {\bf 90},  146402 (2003).

\bibitem{hub}
J. Hubbard,
Proc. R. Soc. London {\bf A276}, 238 (1963); {\bf A281}, 401 (1964);
J. Kanamori,
Prog. Theo. Phys. {\bf 30}, 275 (1963);
M.C. Gutzwiller,
Phys. Rev. Lett. {\bf 10}, 159 (1963).

\bibitem{levy}
M. Levy, Proc. Natl. Acad. Sci. U.S.A. {\bf 76}, 6062 (1979).

\bibitem{ldft4order} 
R. L\'opez-Sandoval and G.M. Pastor, 
Phys. Rev. B {\bf 69}, 085101 (2004). 

\bibitem{nonbip}
For simplicity, we focus here on $\gamma_{ij} > 0$ which is the
relevant case for $t_{ij} = -t < 0$ [see Eq.~(\protect\ref{eq:EK})].
Consequently, we also consider only $\gamma_{ij}^0 > 0$ and 
$\gamma_{ij}^\infty > 0$. However notice that in non-bipartite 
lattices as the C$_{20}$ cluster the domains of representability 
for positive and negative values of the NN $\gamma_{ij}$ are 
different. Therefore, one would have to distinguish the cases
$0\le\gamma_{ij}^{\infty +} \le \gamma_{ij} \le \gamma_{ij}^{0+}$ 
from 
$\gamma_{ij}^{0-} \le \gamma_{ij} \le \gamma_{ij}^{\infty -}\le0$,
since in non-bipartite lattices
$\gamma_{ij}^{0-} \not= -\gamma_{ij}^{0+}$ and
$\gamma_{ij}^{\infty -} \not= -\gamma_{ij}^{\infty +}$.
Nevertheless, as shown in Ref.~\protect\onlinecite{ldftxcfun},
the same scaling behavior of $W / E_{\rm HF}$ as a function of 
$g_{ij} = (\gamma_{ij}      - \gamma_{ij}^{\infty-}) / 
          (\gamma_{ij}^{0-} - \gamma_{ij}^{\infty-})$
is found for 
$\gamma_{ij}^{0-} \le \gamma_{ij} \le \gamma_{ij}^{\infty -}\le0$.
The situation is analogous to what is found in dimerized 
systems where the domain of representability 
$[\gamma^\infty(\phi), \gamma^0(\phi)]$ depends 
on the ratio $\gamma_{12}/\gamma_{23}= \tan\phi$ between 
the bond orders corresponding to short and long bonds
(see Ref.~\protect\onlinecite{ldftdim1D}).  

\bibitem{ldftdim1D}
R. L\'opez-Sandoval and G.M. Pastor, 
Phys. Rev. B {\bf 67}, 035115 (2003).

\bibitem{exact-coef}
The expansion coefficient of the exact functional $W$ are here
obtained from numerical exact diagonalizations for the Heisenberg
limit of the Hubbard model ($\gamma_{ij} \to 0$). For extended systems
or larger clusters they can be calculated by using degenerate 
perturbation theory (see Ref.~\protect\onlinecite{taka}).

\bibitem{taka} 
M. Takahashi, J. Phys. C {\bf 10}, 1289 (1977).  

\bibitem{foot-acc}
A more detailed discussion of the accuracy of LDFT and the 
interaction-energy functionals may be found in 
Refs.~\protect\onlinecite{ldftscfer}, 
\protect\onlinecite{ldft4order}, and \protect\onlinecite{ldftdim1D}, 
where applications to periodic  
one-, two-, and three-dimensional systems are reported.

\bibitem{bois}
L.M.\ Falicov and R.A. Harris, 
J.\ Chem.\ Phys.\ {\bf 51}, 3153 (1969);
S.L.\ Reindl and G.M.\ Pastor,
Phys.\ Rev.\ B {\bf 47}, 4680 (1993).


\end{thebibliography}
\end{document}